\documentclass[aps,prd,groupedaddress,showpacs,nofootinbib]{revtex4}
\usepackage{graphicx,amssymb,bm}

\def\beq{\begin{eqnarray}}    
\def\eeq{\end{eqnarray}}      

\newcommand{\be}{\begin{equation}}
\newcommand{\ee}{\end{equation}}
\newcommand{\bea}{\begin{eqnarray}}
\newcommand{\eea}{\end{eqnarray}}
\newcommand{\beaa}{\begin{eqnarray*}}
\newcommand{\eeaa}{\end{eqnarray*}}

\newcommand{\nn}{\nonumber \\}
\newcommand{\e}{{\rm e}}

\begin{document}



\tolerance=5000

\title{On the way from matter-dominated era to dark energy universe}
\author{Shin'ichi Nojiri}
\email{nojiri@phys.nagoya-u.ac.jp}
\affiliation{Department of Physics, Nagoya University, Nagoya 464-8602. Japan}
\author{Sergei D. Odintsov\footnote{also at Lab. Fundam. Study, Tomsk State
Pedagogical University, Tomsk}}
\email{odintsov@ieec.uab.es}
\affiliation{Instituci\`{o} Catalana de Recerca i Estudis Avan\c{c}ats (ICREA)
and Institut de Ciencies de l'Espai (IEEC-CSIC),
Campus UAB, Facultat de Ciencies, Torre C5-Par-2a pl, E-08193 Bellaterra
(Barcelona), Spain}
\author{Hrvoje \v{S}tefan\v{c}i\'{c} \footnote{On leave of absence from
the Theoretical Physics Division, Rudjer Bo\v{s}kovi\'{c} Institute, Zagreb, Croatia.}}
\email{stefancic@ecm.ub.es, shrvoje@thphys.irb.hr}
\affiliation{Universitat de Barcelona,
Departament d'Estructura i Constituents de la Materia,
Facultat de Fisica, Diagonal 647,
E-08028 Barcelona, Catalonia, Spain}

\begin{abstract}

We develop the general program of the unification of matter-dominated era
with acceleration epoch for scalar-tensor theory or dark fluid.
The general reconstruction of single scalar-tensor theory is fulfilled.
The explicit form of scalar potential for which the theory admits
matter-dominated era, transition to acceleration and (asymptotically
deSitter) acceleration epoch consistent with WMAP data is found.
The interrelation of the epochs of deceleration-acceleration transition
and matter dominance-dark energy transition for dark fluids with general
EOS is investigated. We give several examples of such models with explicit
EOS (using redshift parametrization) where matter-dark energy domination
transition may precede the deceleration-acceleration transition.
As some by-product, the  reconstruction scheme is applied to
scalar-tensor theory to define the scalar potentials which may produce the
dark matter effect. The obtained modification of Newton potential may
explain the rotation curves of galaxies.

\end{abstract}


\maketitle

\section{Introduction}

The dark energy era origin remains to be one of the
challenges of modern cosmology. It has been proposed number of various
models aiming at the description of dark energy universe (for recent
review, see \cite{review}). However, even being consistent with recent
WMAP data, such models turn out to be problematic at the early and
intermediate universe. In other words, we are lacking the unified
  theory which describes at once the sequence of the well-known
cosmological epochs: inflation, radiation/matter dominated epoch and the
current universe speed-up. In such a situation, the attempt to unify
several known cosmological phases may be considered as some approximation
for such a complete theory. From another side, it may rule out number of
dark energy models which being consistent with recent WMAP data cannot
describe correctly the earlier cosmological phases.

In the present paper we develop the general program of the search of dark
energy models (scalar-tensor theory or ideal fluid)  where the sequence of
matter dominated era and acceleration
era is realized. The paper is organized as follows.
Section \ref{sec2} is devoted to the reconstruction program of scalar-tensor
theory admitting matter dominated era before the current acceleration.
In the class of theories with single scalar it is explicitly demonstrated
for which scalar potentials the matter dominated stage precedes the
acceleration era. Clearly, the method may be generalized for multi-scalar
case (compare with \cite{rec}).
In section \ref{sec3} we study the interrelation of the epochs
of deceleration-acceleration transition and the matter-dark energy domination
transition in the class of dark energy fluids with a general EOS. In the
subsection \ref{sec3.1}
we develop the general setting for this class of models,
present some general results and the interesting special cases.
In subsection \ref{sec3.2} we give examples of dark energy fluids,
defined by the concrete EOS, in which in certain parametric regimes the epoch of
the matter-dark energy domination transition may precede the epoch of
the deceleration-acceleration transition.
In subsection \ref{sec3.3} we analyze a redshift parametrization
of the dark energy EOS and elaborate the conditions on the parameters
to observe the matter-dark energy domination transition before
the deceleration-acceleration transition.
In the section \ref{sec4} we show that the same reconstruction method of section
II may be applied to explain the qualitatively different phenomena of dark
matter from scalar-tensor theory
with specific potential. In particulary, it is shown which types of scalar
potential lead to requested modification of Newton potential at large
distances so that rotation curves of galaxies may be explained.
Some summary and outlook are given in the Discussion.

\section{Matter dominated and acceleration era in scalar-tensor theory}

\label{sec2}

In the present section the reconstruction program in
scalar-tensor theory is developed in such a way that the sequence of matter-dominated
and acceleration phases may be achieved for some potentials.
We follow the approach developed in refs.\cite{rec}.
One may begin with the following action of scalar-tensor theory:
\be
\label{STm1}
S=\int d^4 x \sqrt{-g}\left\{\frac{1}{2\kappa^2}R - \frac{1}{2}\omega(\phi)\partial_\mu \phi
\partial^\mu\phi - V(\phi)\right\} + S_m\ .
\ee
Here $\omega(\phi)$ and $V(\phi)$ are proper functions of the scalar field $\phi$ and $S_m$ 
is the action of matter fields. 
We now assume the spatially-flat FRW metric 
\be
\label{FRW}
ds^2 = - dt^2 + a(t)^2 \sum_{i=1}^3 \left(dx^i\right)^2
\ee
Let the scalar field $\phi$ only depends on the time coordinate $t$. Then the FRW
equations are given by 
\be 
\label{any1} 
\frac{3}{\kappa^2}H^2 = \rho + \rho_m\ , \quad - \frac{2}{\kappa^2}\dot H= p + \rho + p_m + \rho_m\ . 
\ee 
Here $\rho_m$ and $p_m$ are the energy density and the pressure of the matter fields respectively. 
The energy density $\rho$ and the pressure $p$ for the scalar field $\phi$ are given by
\be
\label{k4}
\rho = \frac{1}{2}\omega(\phi){\dot \phi}^2 + V(\phi)\ ,\quad p = \frac{1}{2}\omega(\phi){\dot \phi}^2 - V(\phi)\ .
\ee
We also has the scalar field equation given by the variation of the scalar $\phi$:
\be
\label{k8}
0=\omega(\phi)\ddot \phi + \frac{1}{2}\omega'(\phi){\dot\phi}^2 + 3H\omega(\phi)\dot\phi
+ V'(\phi)\ .
\ee

First we consider the case that the matter fields can be neglected by putting $\rho_m=p_m=0$. 
Let $\omega(\phi)$ and $V(\phi)$ are expressed in terms of the single
function $f(\phi)$ \cite{rec} as follows:
\be
\label{STm2}
\omega(\phi)=- \frac{2}{\kappa^2}f'(\phi)\ ,\quad
V(\phi)=\frac{1}{\kappa^2}\left(3f(\phi)^2 + f'(\phi)\right)
\ee
Then we can easily find the following solution for Eqs.(\ref{any1}), (\ref{k4}), and (\ref{k8}) \cite{rec}:
\be
\label{STm3}
\phi=t\ ,\quad H=f(t)\ .
\ee
We should note $\phi$ inside the function $f$ is replace by $t$ since $t$ is associated with $phi$ as in 
the above first equation.
In case that $\omega(\phi)$ is always positive,
the function $\omega$ can be absorbed
into the redefinition of the scalar field
\be
\label{STm4}
\varphi \equiv \int^\phi d\phi \sqrt{\omega(\phi)}
\ee
and the action (\ref{STm1}) has the canonical form:
\be
\label{STm5}
S=\int d^4 x \sqrt{-g}\left\{\frac{1}{2\kappa^2}R - \frac{1}{2}\partial_\mu \varphi \partial^\mu\varphi
  - \tilde V(\varphi)\right\}\ .
\ee
Here the potential $\tilde V(\varphi)$ is defined by
\be
\label{STm6}
\tilde V(\varphi)\equiv V(\phi(\varphi))\ .
\ee

The matter may be included into the action (\ref{STm1}).
Such a matter, in general, interacts with the dark energy.
Then the separation of the total energy density
to the contribution from the dark energy and that from
the matter $\rho_{\rm total}=\rho + \rho_m$ is not unique.
However, if it is defined as
\be
\label{STm7}
p_m\equiv -\rho_m + \frac{\dot \rho}{3H}\ , \quad p\equiv p_{\rm total} - p_m\ ,
\ee
then the matter and the dark energy satisfy the conservation law, separately:
\be
\label{STm8}
\dot \rho_m + 3H\left(\rho_m + p_m\right)=0\ ,\quad
\dot \rho + 3H\left(\rho + p\right)=0\ .
\ee
Especially if $w_m=p_m/\rho_m$ is constant, it follows
\be
\label{STm9}
\rho_m=\rho_{m0} a^{-3(1+w_m)}\ .
\ee
Here $\rho_{m0}$ is a constant.
Hence, for the case that $\omega(\phi)$ and $V(\phi)$ are given by a
single function $g(\phi)$ as follows,
\bea
\label{STm10}
\omega(\phi) &=&- \frac{2}{\kappa^2}g''(\phi) - \left(w_m + 1\right)g_0 \e^{-3(1+w_m)g(\phi)}\ ,\nn
V(\phi) &=& \frac{1}{\kappa^2}\left(3g'(\phi)^2 + g''(\phi)\right) +\frac{w_m -1}{2}g_0 \e^{-3(1+w_m)g(\phi)} \ ,
\eea
we find the solution:
\be
\label{STm11}
\phi=t\ ,\quad H=g'(t)\ ,\quad
\left(a=a_0 \e^{g(t)}\ ,\quad a_0\equiv
\left(\frac{\rho_{m0}}{g_0}\right)^{\frac{1}{3(1+w_m)}}\right)  \ .
\ee

The present universe is expanding with acceleration. On the other hand,
there occured the earlier matter-dominated period,
where the scale factor $a$ behaves as $a\sim t^{2/3}$.
Such behavior could be generated by dust in the
Einstein gravity. The baryons are dust and (cold) dark matter could be a dust.
The ratio of the baryons and
the dark matter in the present universe could be $1:5$ or $1:6$,
which should not be changed even in matter dominant
era. It is not clear what the dark matter is. For instance, the dark
matter might not be the real matter but some
(effective) artifact which appears in the modified/scalar-tensor gravity.
In the present paper, it is assumed that not only dark energy
but also the dark matter originates from the scalar field $\phi$.

We now investigate that the transition from the matter dominant
period to the acceleration period could be realized in the present
formulation. In the following, the
contribution from matter is neglected since the ratio of the matter with
the (effective) dark matter could be small.

First example is
\be
\label{STm12}
H=f(t)=g_0 + \frac{g_1}{t}\ .
\ee
When $t$ is large, the first term in (\ref{STm12}) dominates and
the Hubble rate $H$ becomes a constant.
Therefore, the universe is asymptotically deSitter space, which is an
accelerating universe (for recent examples of late-time accelerating cosmology
in scalar-tensor theory, see \cite{faraoni,odnots} and for earlier study 
and list of refs. see \cite{maeda,barrow}). On the
other hand, when $t$ is small, the second term in (\ref{STm12}) dominates
and the scale factor behaves as $a\sim t^{g_1}$. Therefore if $g_1=2/3$,
the matter-dominated period could be realized.
By substituting (\ref{STm12}) into (\ref{STm2}), one finds
\be
\label{STm13}
\omega(\phi) = \frac{2}{\kappa^2}\frac{g_1}{\phi^2}\ ,\quad
V(\phi) = \frac{1}{\kappa^2}\left(3g_0^2 + \frac{6g_0 g_1}{\phi} + \frac{3g_1^2 - g_1}{\phi^2}\right)\ .
\ee
Hence, (\ref{STm4}) shows
\be
\label{STm14}
\varphi = \frac{\sqrt{2g_1}}{\kappa \ln \frac{\phi}{\phi_0}}\ .
\ee
Here $\phi_0$ is a constant. In terms of the canonical field $\varphi$,
the potential $\tilde V(\varphi)$
(\ref{STm6}) is given by
\be
\label{STm15}
\tilde V(\varphi) = \frac{1}{\kappa^2}\left(3g_0^2 + \frac{6g_0g_1}{\phi_0}\e^{-\kappa\varphi/\sqrt{2g_1}}
+ \frac{3g_1^2 - g_1}{\phi_0^2}\e^{-2\kappa\varphi/\sqrt{2g_1}} \right)\ .
\ee

Before going to the second example,
we consider the Einstein gravity with cosmological constant and with
matter characterized by the EOS parameter $w$.
FRW equation has the following form:
\be
\label{LCDM1}
\frac{3}{\kappa^2}H^2 = \rho_0 a^{-3(1+w)} + \frac{3}{\kappa^2 l^2}\ .
\ee
Here $l$ is the length parameter coming from the cosmological constant.
The solution of (\ref{LCDM1}) is given by
\bea
\label{LCDM2}
a&=&a_0\e^{g(t)}\ ,\nn
g(t)&=&\frac{2}{3(1+w)}\ln \left(\alpha \sinh \left(\frac{3(1+w)}{2l}\left(t - t_0 \right)\right)\right)\ .
\eea
Here $t_0$ is a constant of the integration and
\be
\label{LCDM3}
\alpha^2\equiv \frac{1}{3}\kappa^2 l^2 \rho_0 a_0^{-3(1+w)}\ .
\ee
As an second example, we consider (\ref{LCDM2}) without matter.
In this case, (\ref{STm2}) shows
\bea
\label{STm16}
&& \omega(\phi)= \frac{3(1+w)}{\kappa^2 l^2}\sinh^{-2}\left(\frac{3(1+w)}{2l}\left(\phi - t_0\right)\right)\ ,\nn
&& V(\phi)= \frac{1}{\kappa^2}\left(\frac{3}{l^2}\coth^2\left(\frac{3(1+w)}{2l}\left(\phi - t_0\right)\right)
  - \frac{3(1+w)}{2l^2}\sinh^{-2}\left(\frac{3(1+w)}{2l}\left(\phi - t_0\right)\right)\right)\ .
\eea
Eq.(\ref{STm4}) and (\ref{STm6}) also indicate
\bea
\label{STm17}
\varphi&=&\frac{2}{\kappa\sqrt{3(1+w)}} \ln \tanh \left(\frac{3(1+w)}{4l}\left(\phi - t_0\right)\right)\ , \nn
V(\varphi)&=& \frac{1}{\kappa^2}\left(\frac{3}{l^2}\cosh^2\left(\frac{\varphi}{\varphi_0}\right)
  - \frac{3(1+w)}{2l^2}\sinh^2\left(\frac{\varphi}{\varphi_0}\right)\right)\ ,\nn
\frac{1}{\varphi_0}&\equiv & \frac{\kappa}{2}\sqrt{\frac{1+w}{\alpha}}\ .
\eea
Thus, in both examples, (\ref{STm12}) and (\ref{LCDM2}),
there occurs the transition from
matter dominated phase to acceleration phase. In the acceleration phase,
in the above examples, the universe asymptotically
approaches to deSitter space. This does not conflict with WMAP data.
Indeed,
three years WMAP data have been analyzed in Ref.\cite{Spergel}.
The combined analysis of WMAP with supernova Legacy
survey (SNLS) constrains the dark energy equation of state $w_{DE}$ pushing it
towards the cosmological constant. The marginalized best fit values of the
equation of state parameter at 68$\%$ confidance level
are given by $-1.14\leq w_{\rm DE} \leq -0.93$. In case of a prior that universe is
flat, the combined data gives $-1.06 \leq w_{\rm DE} \leq -0.90 $.
In the examples (\ref{STm12}) and (\ref{LCDM2}),
the universe goes to asymptotically deSitter space, which gives
$w_{\rm DE}\to -1$, which does not, of course, conflict
with the above constraints.
Note, however, one needs to fine-tune $g_0$ in (\ref{STm12})
and $1/l$ in (\ref{LCDM2}) to be
$g_0\sim 1/l \sim 10^{-33}$ eV, in order to reproduce
  the observed Hubble rate
$H_0\sim 70$ km$\,$s$^{-1}$Mpc$^{-1}\sim 10^{-33}$ eV.

In principle, more complicated examples of multi-scalar-tensor theory
reconstruction may be considered. These (radiation/matter dominated)
regimes may be related with dark energy stages proposed in \cite{rec}.
In the next section, it will be demonstrated that similar scenario may be
realized for dark fluids.

Before ending this section, we consider what could happen in the Jordan frame.
By transforming the metric by a function $\sigma(\phi)$ of the scalar field $\phi$ as
$g_{\mu\nu} \to \e^{\sigma(\phi)} g_{\mu\nu}$, the action (\ref{STm1}) is transformed as
\be
\label{BD1}
S \to S_J=\int d^4 x \sqrt{-g}\e^{\sigma(\phi)} \left\{\frac{1}{2\kappa^2}R 
 - \left(\frac{1}{2}\omega(\phi) - \frac{3}{2}{\sigma'(\phi)}^2\right)\partial_\mu \phi
\partial^\mu\phi - \e^{\sigma(\phi)}V(\phi)\right\}\ .
\ee
Especially if we consider the case $\sigma(\phi)=2\phi/\phi_0$ with a positive constant $\phi_0$, we find 
\be
\label{BD2}
S_J=\int d^4 x \sqrt{-g}\e^{2\phi/\phi_0} \left\{\frac{1}{2\kappa^2}R 
 - \left(\frac{1}{2}\omega(\phi) - \frac{6}{\phi_0^2}\right)\partial_\mu \phi
\partial^\mu\phi - \e^{2\phi/\phi_0}V(\phi)\right\}\ .
\ee
Then we obtain the Jordan frame action. 

We should note, however, that the solution is the Jordan frame does not corresponds to the real 
cosmology. As an example, we consider the case (\ref{STm12}) by choosing $\phi=t$. 
Then we have the following relations between the quantities in the original Einstein frame and 
those in the Jordan frame, which are distinguished by the suffix $J$: 
\be
\label{BD3}
a_J=\e^{\phi/\phi_0} a\ ,\quad t_J=\phi_0 \e^{t/\phi_0}
\ee
Then the Hubble rate in the Jordan frame is given by
\be
\label{BD4}
H_J=\frac{1}{a_J}\frac{da_J}{dt_J}=\frac{1 + \phi_0 g_0 + \frac{g_1}{\ln\frac{t_J}{\phi_0}} }{t_J}\ . 
\ee
Different from the case of (\ref{STm12}), the solution does not go to asymptotically deSitter space in the 
limit of $t_J\to +\infty$,
\be
\label{BD5}
H_J\to \frac{1 + \phi_0 g_0 }{t_J}\ . 
\ee
Then the cosmology looks to change in the frame. 
If we neglect the matter, we cannot say which frame should be physical. But if we couple the matter, in the original 
Einstein frame (\ref{STm1}), the matter does not couple with the scalar field $\phi$ but in the Jordan frame, the matter 
couples with $\phi$ through the scale transformation $g_{\mu\nu} \to \e^{2\phi/\phi_0} g_{\mu\nu}$. Then in the present 
model, the Jordan frame would not be physical.

\section{Matter-dark energy domination transition versus deceleration-acceleration transition}

\label{sec3}

\subsection{The general setting}

\label{sec3.1}

In the present section,
we consider a wide class of cosmological models to study the interrelation
of two epochs, the epoch of the transition
from the matter dominated to the dark energy dominated universe
and the epoch of the deceleration-acceleration transition.
Unlike to previous section, we limit ourselves by the consideration of the
FRW universe filled with dark energy ideal fluid of general form.
We assume that the universe is spatially flat and
that it contains two components, a dark energy component, and a matter
component, which is presently nonrelativistic.
In the study of this class of models we are especially interested in the
possibility that the parameter of dark energy EOS may be variable with redshift
and that dark energy at high redshifts may have different properties than at
low redshifts (i.e. that at higher redshifts it could even be matter-like).
The dependence of the nonrelativistic matter component on the redshift
(scale factor) is the standard one,
$\rho_m(z)=\rho_{m,0} (1+z)^3$ $(\rho_m(a)=\rho_{m,0} (a/a_0)^{-3}$),
whereas the behavior of the dark energy component
is defined by its equation of state (EOS):
\begin{equation}
\label{eq:defeos}
p_d(z)=w_d(z) \rho_d(z) \, ,
\end{equation}
\begin{equation}
\label{eq:defeos2}
p_d(a)=w_d(a) \rho_d(a) \, , \nonumber
\end{equation}
where $w_d(z)$ $(w_d(a))$ is in general a function of redshift (scale factor),
  i.e. we discuss a very general class of
dark energy models. The dark energy density can be written in the form
\begin{equation}
\label{eq:dyn}
\rho_d(z)=\rho_{d,0} \exp \left( 3 \int_0^z \frac{1+w_d(z')}{1+z'}d z' \right) =
\rho_{d,0} (1+z)^3 \exp \left( 3 \int_0^z \frac{w_d(z')}{1+z'}d z' \right) \, .
\end{equation}
\begin{equation}
\label{eq:dyn2}
\rho_d(a)=\rho_{d,0} \exp \left(- 3 \int_{a_0}^a \frac{1+w_d(a')}{a'}d a' \right) =
\rho_{d,0} \left(\frac{a}{a_0} \right)^{-3} \exp \left( -3 \int_{a_0}^a
\frac{w_d(a')}{a'}d a' \right) \, . \nonumber
\end{equation}
This expression allows us to write the ratio
  $r(z)=\rho_d(z)/\rho_m(z)$ ($r(a)=\rho_d(a)/\rho_m(a)$) as
\begin{equation}
\label{eq:ratio}
r(z)=\frac{1-\Omega_m^0}{\Omega_m^0} \exp \left(3 \int_0^z \frac{w_d(z')}{1+z'}d z' \right) \, .
\end{equation}
\begin{equation}
\label{eq:ratio2}
r(a)=\frac{1-\Omega_m^0}{\Omega_m^0} \exp \left(-3 \int_{a_0}^a \frac{w_d(a')}{a'}d a' \right) \, . \nonumber
\end{equation}
The description of the cosmological evolution is completed with the Hubble equation
\begin{equation}
\label{eq:hubble}
H^2=\frac{8 \pi G}{3} \rho_m (1 + r) \,
\end{equation}
and the expression for the acceleration of the expansion of the universe
\begin{equation}
\label{eq:acc}
\frac{\ddot{a}}{a}= -\frac{4 \pi G}{3} \rho_m (1 + r(1+3 w_d)) \, .
\end{equation}

The main objects of our present study are two redshifts (scale factor values),
  the redshift $z_{EQ}$ (scale factor $a_{EQ}$)
at which the matter dominated epoch ends and the dark energy dominated epoch
  begins, and the redshift $z_{D/A}$ (scale factor
$a_{D/A}$) at which the universe transits from the decelerated expansion to the accelerated one.
The redshift (scale factor) of the dark energy domination onset $z_{EQ}$ ($a_{EQ}$) is characterized by $r(z_{EQ})=1$
($r(a_{EQ})=1$). This requirement can be further elaborated
  using (\ref{eq:ratio}):
\begin{equation}
\label{eq:eq}
r(z_{EQ})=\frac{1-\Omega_m^0}{\Omega_m^0} \exp \left(3 \int_0^{z_{EQ}} \frac{w_d(z')}{1+z'}d z' \right) = 1\, .
\end{equation}
\begin{equation}
\label{eq:eq2}
r(a_{EQ})=\frac{1-\Omega_m^0}{\Omega_m^0} \exp \left(-3 \int_{a_0}^{a_{EQ}} \frac{w_d(a')}{a'}d a' \right) = 1\, . \nonumber
\end{equation}
The deceleration-acceleration transition redshift $z_{D/A}$ (scale factor $a_{D/A}$) is the point where the acceleration of
the expansion of the universe vanishes which leads to
\begin{equation}
\label{eq:da}
1+ (1+ 3 w_d(z_{D/A})) \frac{1-\Omega_m^0}{\Omega_m^0} \exp \left( 3 \int_0^{z_{D/A}} \frac{w_d(z')}{1+z'}d z' \right) = 0 \, .
\end{equation}
\begin{equation}
\label{eq:da2}
1+ (1+ 3 w_d(a_{D/A})) \frac{1-\Omega_m^0}{\Omega_m^0} \exp \left( -3 \int_{a_0}^{a_{D/A}} \frac{w_d(a')}{a'}d a' \right) = 0 \, . \nonumber
\end{equation}
The expressions (\ref{eq:eq}) and (\ref{eq:da}) can be conveniently combined into a compact expression depending only
on the evolution of the function $w_d(z)$ ($w_d(a)$) in the interval between $z_{EQ}$ and $z_{D/A}$ ($a_{EQ}$ and $a_{D/A}$):
\begin{equation}
\label{eq:gen}
\exp \left(3 \int_{z_{D/A}}^{z_{EQ}} \frac{w_d(z')}{1+z'}d z' \right) = -(1+3 w_d(z_{D/A})) \, .
\end{equation}
\begin{equation}
\label{eq:gen2}
\exp \left(-3 \int_{a_{D/A}}^{a_{EQ}} \frac{w_d(a')}{a'}d a' \right) = -(1+3 w_d(a_{D/A})) \, . \nonumber
\end{equation}
This expression, gives a (trivial) condition $w_d(z_{D/A})<-1/3$ ($w_d(a_{D/A})<-1/3$).

All results obtained so far do not depend on the ordering of $z_{EQ}$ and $z_{D/A}$ ($a_{EQ}$ and $a_{D/A}$).
For a large class of dark energy models, including the benchmark $\Lambda$CDM model, the deceleration-acceleration
transition happens before the onset of the dark-energy dominated epoch. Here we focus our attention on an alternative
scenario in which the transition from the matter domination to the dark energy domination precedes
the deceleration-acceleration transition, i.e. $z_{EQ} \ge z_{D/A}$ ($a_{EQ} \le a_{D/A}$). Physically, this ordering of the aforementioned epochs suggests that the dark energy component may acquire matter-like properties at higher redshifts, i.e. might be dominant component but still nonaccelerating in character and only at the small redshifts become truly accelerating component, as indicated in section \ref{sec2}.

For this scenario, using the expressions given above, further general information can be obtained. At $z_{EQ}$ ($a_{EQ}$)
the expansion of the universe is decelerated. From (\ref{eq:acc}) one obtains the condition
$1 + r(z_{EQ})(1+3 w_d(z_{EQ})) \ge 0$ ($1 + r(a_{EQ})(1+3 w_d(a_{EQ})) \ge 0$), which, using the fact
that $r(z_{EQ})=1$ ($r(a_{EQ})=1$) leads to the condition $w_d(z_{EQ}) \ge -2/3$ ($w_d(a_{EQ}) \ge -2/3$).
On the other hand, for $z > z_{EQ}$ ($a < a_{EQ}$) the universe is matter dominated and $r < 1$.
Therefore, at $z_{EQ}$ ($a_{EQ}$), $r(z)$ is a decreasing function of redshift ($r(a)$ is an increasing function of scale factor).
This fact can be further expressed as
\begin{equation}
\label{eq:difeq}
\left. \frac{d r}{d z} \right|_{z=z_{EQ}} = 3 \frac{w_d(z_{EQ})}{1+z_{EQ}} < 0 \, ,
\end{equation}
\begin{equation}
\label{eq:difeq2}
\left. \frac{d r}{d a} \right|_{a=a_{EQ}} = - 3 \frac{w_d(a_{EQ})}{a_{EQ}} > 0 \, , \nonumber
\end{equation}
which leads to the condition $w_d(z_{EQ}) < 0$ ($w_d(a_{EQ}) < 0$). Combining this result with the one coming from
the deceleration of the universe at $z_{EQ}$ ($a_{EQ}$) we finally obtain $-2/3 \le w_d(z_{EQ}) < 0$ ($-2/3 \le w_d(a_{EQ}) < 0$).
Since at $z_{D/A}$ ($a_{D/A}$) the condition $w_d(z_{D/A})<-1/3$ ($w_d(a_{D/A})<-1/3$) must be satisfied, for models with
the monotonous variation of $w_d(z)$ ($w_d(a)$) in the $(z_{EQ}, z_{D/A})$ interval ($(a_{EQ}, a_{D/A})$ interval), it follows
that $w_d(z)$ ($w_d(a)$) is negative in the same interval. The expression
\begin{equation}
\label{eq:racc}
r(z_{D/A})= \exp \left(-3 \int_{z_{D/A}}^{z_{EQ}} \frac{w_d(z')}{1+z'}d z' \right) \, ,
\end{equation}
\begin{equation}
\label{eq:racc2}
r(a_{D/A})= \exp \left(-3 \int_{a_{EQ}}^{a_{D/A}} \frac{w_d(a')}{a'}d a' \right) \, , \nonumber
\end{equation}
which immediately leads to $r(z_{D/A}) > 1$ ($r(a_{D/A}) > 1$).

An interesting limiting case is the one when $z_{EQ}=z_{D/A}$ ($a_{EQ}=a_{D/A}$). From (\ref{eq:gen}) it is straightforward
to obtain $w_d(z_{EQ}=z_{D/A})=-2/3$ ($w_d(a_{EQ}=a_{D/A})=-2/3$). In general, the redshift dependence of $w_d(z)$ ($w_d(a)$)
will contain a number of parameters. This condition on $w_d(z_{EQ}=z_{D/A})$ ($w_d(a_{EQ}=a_{D/A})$) determines the boundary
of the part of the parametric space where the studied scenario of the expansion of the universe is realized.

Finally, let us study the simplest case when $w_d(z)=w_d=const$ ($w_d(a)=w_d=const$). In this case we have
\begin{equation}
\label{eq:wconst}
\left( \frac{1+z_{EQ}}{1+z_{D/A}} \right)^{ 3 w_d} = -(1+3 w_d) \, .
\end{equation}
\begin{equation}
\label{eq:wconst2}
\left( \frac{a_{D/A}}{a_{EQ}} \right)^{ 3 w_d} = -(1+3 w_d) \, . \nonumber
\end{equation}
Since in the studied scenario we have $z_{EQ} \ge z_{D/A}$ ($a_{EQ} \le a_{D/A}$), the relation (\ref{eq:wconst}) gives
a constraint $w_d \ge -2/3$. In the limiting case $z_{EQ}=z_{D/A}$ ($a_{EQ}=a_{D/A}$) we obtain $w_d = -2/3$, in accordance
with the general result obtained in the previous paragraph.

\subsection{Examples of dark energy models}

\label{sec3.2}

We further discuss two concrete examples of the dark energy models which result in the studied scenario. Let us first discuss
the dark energy model, first introduced in \cite{odno} and
then analyzed in detail in \cite{stef} and \cite{odnots}, defined
by the EOS
\begin{equation}
\label{eq:eos1}
p_d=-\rho_d-A \rho_d^{\alpha} \, .
\end{equation}
This model exhibits a rich variety of interesting phenomena
\cite{odno,stef,odnots} in different parameter regimes, including
the cosmological singularity at the finite value of the scale factor,
  Big Rip-like singularities \cite{brett}
(for a classification of future cosmological singularities see \cite{odnots}).
  The dependence of the dark energy density on
scale factor can be given as a simple analytic expression
\begin{equation}
\label{eq:rhodex1}
\rho_d=\rho_{d,0} \left( 1 + 3 \tilde{A} (1- \alpha) \ln \frac{a}{a_0} \right)^{1/(1-\alpha)} \, ,
\end{equation}
where $\tilde{A}=A \rho_d^{\alpha-1}$. Moreover, in the situation where the matter density or the spatial curvature term
in the Hubble equation are negligible compared to the dark energy density, an analytic expression for the
time evolution of the scale factor can be obtained \cite{stef}. For $\alpha \neq 1/2$ we have
\be
\label{eq:aodt}
\left( 1 + 3 \tilde{A} (1-\alpha) \ln \frac{a_{1}}{a_{0}} \right)^{\frac{1-2\alpha}{2(1-\alpha)}}
- \left( 1 + 3 \tilde{A} (1-\alpha) \ln \frac{a_{2}}{a_{0}} \right)^{\frac{1-2\alpha}{2(1-\alpha)}} 
=\frac{3}{2} \tilde{A} (1-2\alpha) \Omega_{d,0}^{1/2} H_{0} (t_{1}-t_{2}) \, .
\ee
whereas for $\alpha = 1/2$ the following expression is valid
\begin{equation}
\label{eq:aodtspec}
\ln \frac{1 + \frac{3}{2} \tilde{A} \ln \frac{a_{1}}{a_{0}}}
{1 + \frac{3}{2} \tilde{A} \ln \frac{a_{2}}{a_{0}}} = \frac{3}{2} \tilde{A} \Omega_{d,0}^{1/2} H_{0} (t_{1}-t_{2}) \, .
\end{equation}
The scenario $z_{EQ} > z_{D/A}$ ($a_{EQ} < a_{D/A}$) can be realized for the values $\alpha > 1$ and
$\tilde{A}=A \rho_{d,0}^{\alpha} < 0$. For instance, for $\alpha=2$ and $\tilde{A}=-0.27$ one obtains
$z_{EQ}=0.52$ ($a_{EQ}/a_0=0.66$) and $z_{D/A}=0.42$ ($a_{D/A}/a_0=0.70$). However, in the parameter regime
$\alpha > 1$ and $\tilde{A} < 0$ the dark energy model (\ref{eq:eos1}) is burdened with a following problem:
for redshifts larger than some limiting redshift $z_l$ (scale factor values smaller than some limiting scale value $a_l$)
the dark energy is no longer a well defined function of redshift (scale factor) (for the concrete values of $\alpha=2$
and $\tilde{A}=-0.27$ we have $z_l=2.44$ ($a_l/a_0=0.29$). Clearly, the realization of the studied scenario
in the dark energy model (\ref{eq:eos1}) is physically acceptable only if the description given by (\ref{eq:eos1})
is valid up to some redshift $z_{\mathrm{lim}} < z_l$ (for values of the scale factor bigger than $a_{\mathrm{lim}} > a_l$).
One theoretical possibility might be that the dark energy component (\ref{eq:eos1}) suddenly appears at the redshift
$z_{\mathrm{lim}}$ (the scale value $a_{\mathrm{lim}}$), i.e. that it is negligible before that epoch.

Next we turn to the dark energy model with the following scaling of the dark energy density with redshift
\begin{equation}
\label{eq:eos2}
\rho_{d}=\rho_{d,0} \left[ \frac{(1+z)^{3 (1+w_{1})/b} + C_2 (1+z)^{3 (1+w_{2})/b}}{1+ C_2} \right]^b \, ,
\end{equation}
\begin{equation}
\label{eq:eos2-2}
\rho_{d}=\rho_{d,0} \left[ \frac{\left(\frac{a}{a_0} \right)^{-3 (1+w_{1})/b} + C_2 \left(\frac{a}{a_0}
\right)^{-3 (1+w_{2})/b}}{1+ C_2} \right]^b \, , \nonumber
\end{equation}
where $C_2=(w_{d,0}-w_1)/(w_2-w_{d,0})$. This dark energy model has an implicitly defined equation of state and
it can also exhibit the phenomenon of the cosmological constant boundary crossing. The studied scenario may be
realized in this dark energy model. For illustration purposes, for $w_1=-0.2$, $w_2=-1.3$, $w_{d,0}=-1.05$ and
$b=0.6$ we obtain $z_{EQ}=0.41$ ($a_{EQ}/a_0=0.71$), $z_{D/A}=0.35$ ($a_{D/A}/a_0=0.74$) and the CC boundary crossing
at $z_{cross}=0.045$ ($a_{cross}/a_0=0.96$).

The first example demonstrates how the studied scenario can be realized in a dark energy model with an explicitly
defined equation of state of the type $p_d=-\rho_d -f(\rho_d)$, whereas the second example demonstrates that
the behavior of interest can also be obtained in the dark energy models
with the implicitly defined EOS of the type $F(p_d, \rho_d)=0$ \cite{impl1,impl2,impl3} (the EOS
may contain also inhomogeneous terms \cite{impl2,impl3,brevik}).
These facts show that the above scenario
can be realized in different classes of dark energy models (even in
modified gravity as explicitly shown in \cite{mod}) as already studied in
the general framework of the previous subsection.

\subsection{Example of a dark energy parametrization}

\label{sec3.3}

We further consider a parametrization of the redshift dependence which has recently attracted a lot of attention
in the analysis of the observational data \cite{pad,zhang}
\begin{equation}
\label{eq:par}
w_d(z)=w_0 + w_{0}' \frac{z}{1+z} \, .
\end{equation}

\begin{figure}
\centerline{\resizebox{1.0\textwidth}{!}{\includegraphics{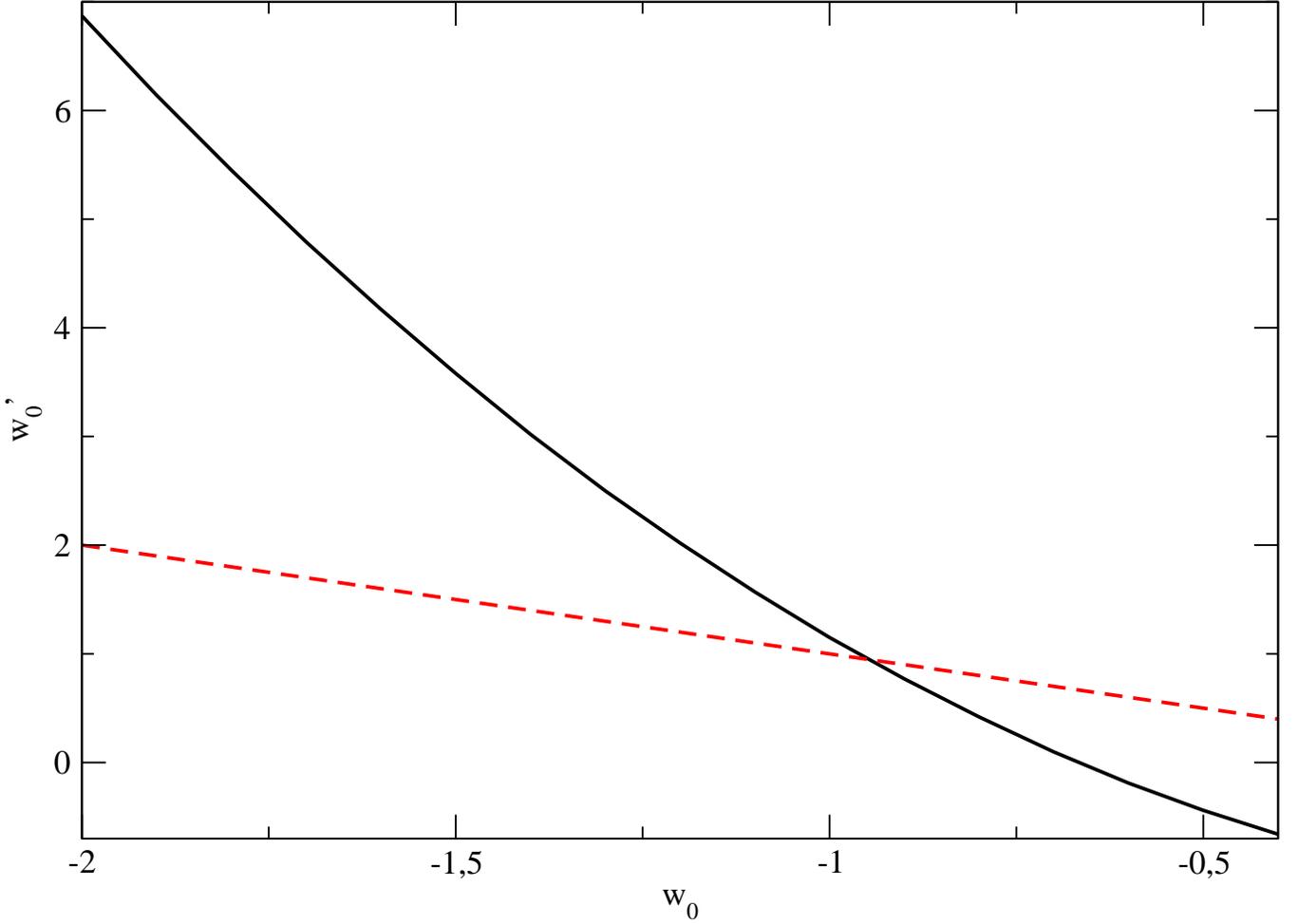}}}
\caption{\label{fig:boundary} The regions with $z_{EQ}>z_{D/A}$ (above the full line curve) and $w_0+w_{0}'<0$
(below the dashed line curve) for $\Omega_m^0=0.3$. }
\end{figure}

To find the part of the parametric space where our scenario is realized we determine the boundary of that part from
the condition $z_{EQ}=z_{D/A}$. We obtain
\begin{equation}
\label{eq:zeqnum}
z_{EQ}=-\frac{2/3+w_0}{2/3+w_0+w_{0}'} \, ,
\end{equation}
which in combination with (\ref{eq:eq}) and (\ref{eq:par}) leads to the condition on $w_0$ and $w_{0}'$
\begin{equation}
\label{eq:condpadpar}
\frac{1-\Omega_m^0}{\Omega_m^0} \left( \frac{w_{0}'}{2/3+w_0+w_{0}'} \right)^{3 (w_0+w_{0}')} e^{3 w_0 +2}=1 \, .
\end{equation}
Solving this equation numerically we obtain the boundary of the part of the parametric space where the studied
scenario is realized. This boundary is depicted in Fig. \ref{fig:boundary}. It is also possible to impose
the condition $w_0+w_{0}'<0$ to obtain the part of the parametric space where our scenario is possible and
the dark energy does not dominate at high redshifts, see Fig. \ref{fig:boundary}.
Thus, the principal possibility of sequence:matter-dominated and
acceleration era is demonstrated for dark energy ideal fluid.

\section{The reconstruction of dark matter from scalar-tensor theory}

\label{sec4}

The existence of the dark matter (for recent review from modified
gravity point of view, see \cite{moffat}) could be
conjectured from the rotation curves of the galaxies and the formation of galaxy
clusters. This means, if the dark matter is not some strange matter, but, for instance,
modified gravity/scalar-tensor theory produces the dark matter, the Newton law should be
modified at large (astrophysical) scales.
Here as an example, we consider scalar-tensor gravity (\ref{STm1})
or (\ref{STm5}). It will be shown (by analogy with the method of second
section) that for some scalar-tensor gravity
with specific potentials the requested modification of Newton law could be achieved.

For the action (\ref{STm1}), the Einstein equation has the following form:
\be
\label{DMst1}
0=\frac{1}{2}\left\{\frac{R}{2\kappa^2} - \frac{1}{2}\omega(\phi)\partial_\rho \phi \partial^\rho \phi
  - V(\phi)\right\}g_{\mu\nu} - \frac{1}{2\kappa^2}R_{\mu\nu} + \frac{1}{2}\omega(\phi)\partial_\mu \phi \partial_\nu \phi\ ,
\ee
and the field equation is:
\be
\label{DMst2}
0=\frac{1}{\sqrt{-g}}\partial_\mu \left(\sqrt{-g}\omega(\phi)\partial^\mu\phi\right)
  - \frac{1}{2}\omega'(\phi) \partial_\rho \phi \partial^\rho \phi - V'(\phi)\ .
\ee
We now assume 4-dimensional spherically symmetric and static metric:
\be
\label{DMst3}
ds^2 = \e^{2\nu(r)}dt^2 + \e^{2\lambda(r)}dr^2 + r^2 \sum_{i,j=1,2} \tilde g_{ij} dx^i dx^j\ .
\ee
Here $\tilde g_{ij}$ is the metric of the unit sphere and $\phi$ only depends on $r$.
Then the $(t,t)$, $(r,r)$, and $(i,j)$ components of
the Einstein equation (\ref{DMst1}) and the field equation (\ref{DMst2})
have the following forms:
\bea
\label{DMst4}
0&=& \frac{\e^{2(\nu - \lambda)}}{2\kappa^2}\left( - \frac{2\lambda'}{r} + \frac{\e^{2\lambda} - 1}{r^2}\right)
+ \frac{1}{4}\omega(\phi)\left(\phi'\right)^2 \e^{2(\nu - \lambda)} + \frac{1}{2}V(\phi) \e^{2\nu}\ ,\\
\label{DMst5}
0&=& \frac{1}{2\kappa^2}\left( \frac{2\nu'}{r} + \frac{\e^{2\lambda} - 1}{r^2}\right)
+ \frac{1}{4}\omega(\phi)\left(\phi'\right)^2 - \frac{1}{2}V(\phi) \e^{2\lambda}\ ,\\
\label{DMst6}
0&=&\frac{\e^{- 2\lambda}r^2}{2\kappa^2}\left( - \nu'' - \left(\nu' - \lambda'\right)\nu' - \frac{\nu' - \lambda'}{r}
+ \frac{2\left(\e^{2\lambda} - 1\right)}{r^2}\right) \nn
&& + r^2\left( - \frac{1}{4}\omega(\phi)\left(\phi'\right)^2 \e^{- 2 \lambda} - \frac{1}{2}V ( \phi ) \right) \ ,\\
\label{DMst7}
0&=&\e^{-2\lambda}\left\{ \omega(\phi) \phi'' + \frac{1}{2}\frac{d\omega(\phi)}{d\phi} \left(\phi'\right)^2
+ \frac{2\omega(\phi)}{r}\phi' - 2\lambda' \omega(\phi)\phi'\right] - \frac{dV(\phi)}{d\phi}\ .
\eea
By using the ambiguity of the redefinition of the scalar field $\phi$,
one may identify $\phi$ with the radial coordinate $r$:
\be
\label{DMst8}
\phi=r\ .
\ee
We should note that this is mathematical trick. The scalar field should not be always identified with the radial 
coordinate $r$ but we should note that there is an ambiguity of the redefinition of $\phi$ like $\phi\to\tilde\phi=\Phi(\phi)$ 
by a proper function $\Phi$. If we redefine $\omega(\phi)$ and $\tilde V(\tilde\phi)$ by 
$\omega(\phi)\to \tilde(\tilde\phi)\equiv \left(\Phi'\left(\phi(\tilde\phi)\right)\right)^2\omega\left(\phi(\tilde\phi)\right)$ 
and $\tilde V(\tilde\phi) \equiv V\left(\phi(\tilde\phi)\right)$, the form of the action (\ref{STm1}) is invariant. 
Then if the scalar field $\phi$ is not constant, at least locally, we may choose to identify the scalar field with $r$. 

By choosing $\phi=r$, Eqs.(\ref{DMst4}-\ref{DMst7}) reduce to
\bea
\label{DMst9}
0&=& \frac{1}{2\kappa^2}\left( - \frac{2\lambda'}{r} + \frac{\e^{2\lambda} - 1}{r^2}\right)
+ \frac{1}{4}\omega(\phi) + \frac{1}{2}V(\phi) \e^{2\lambda}\ ,\\
\label{DMst10}
0&=& \frac{1}{2\kappa^2}\left( \frac{2\nu'}{r} + \frac{\e^{2\lambda} - 1}{r^2}\right)
+ \frac{1}{4}\omega(\phi) - \frac{1}{2}V(\phi) \e^{2\lambda}\ ,\\
\label{DMst11}
0&=&\frac{1}{2\kappa^2}\left( - \nu'' - \left(\nu' - \lambda'\right)\nu' - \frac{\nu' - \lambda'}{r}
+ \frac{2\left(\e^{2\lambda} - 1\right)}{r^2}\right) - \frac{1}{4}\omega(\phi) - \frac{1}{2}V(\phi)\e^{2\lambda} \ ,\\
\label{DMst12}
0&=& \frac{1}{2}\frac{d\omega(\phi)}{d\phi} + \frac{2\omega(\phi)}{r} - 2\lambda' \omega(\phi)
  - \e^{2\lambda}\frac{dV(\phi)}{d\phi}\ .
\eea
Eqs.(\ref{DMst9}-\ref{DMst11}) give
\bea
\label{DMst13}
&& 0= \nu'' + \left(\nu' - \lambda'\right)\nu' + \frac{\nu' + \lambda'}{r}
  - \frac{3\left(\e^{2\lambda} - 1\right)}{r^2} \\
\label{DMst14}
&& \omega(\phi=r) = - \frac{4}{\kappa^2}\left\{\frac{\nu' - \lambda'}{r} + \frac{\e^{2\lambda} - 1}{r^2}\right\}\ ,\\
\label{DMst15}
&& V(\phi=r) = \frac{\e^{-2\lambda}\left(\nu' + \lambda'\right)}{\kappa^2 r}\ .
\eea
Let assume there could be some desired form of $\nu$ ($\lambda$).
Solving (\ref{DMst13}), one finds the form of $\lambda$ ($\nu$).
Then by using Eqs.(\ref{DMst14}) and
(\ref{DMst15}), we get the explicit form of $\omega(\phi)$ and $V(\phi)$,
which generate the requested form of $\mu$ ($\lambda$)
as a solution, as follows:
\bea
\label{DMst16}
\omega(\phi)&=& - \frac{4}{\kappa^2}\left\{\frac{\nu'(\phi) - \lambda'(\phi)}{\phi}
+ \frac{\e^{2\lambda(\phi)} - 1}{\phi^2}\right\}\ ,\\
\label{DMst17}
V(\phi)&=& \frac{\e^{-2\lambda(\phi)}\left(\nu'(\phi) + \lambda'(\phi)\right)}{\kappa^2 \phi}\ .
\eea

As an example, we consider the case
\be
\label{DMst17a}
\e^{-2\lambda}=1 - \frac{M}{r} + \alpha r^{-\beta}\ .
\ee
where it is chosen $0<\beta<1$. If $r$ is small, the last term can be
neglected but if $r$ is large, the last term
dominates if compared with the second term. In the limit $r\to \infty$,
$\e^{-\lambda}\to 1$, which is necessary for the spacetime
to be asymptotically flat.
Let us consider the region where $r$ is large. One can approximate
(\ref{DMst17}) as
\be
\label{DMst17b}
\e^{-2\lambda} \sim 1 + \alpha r^{-\beta}\ .
\ee
Solving (\ref{DMst13}), we find
\be
\label{DMst18}
\e^{2\nu}=1 - \frac{\alpha\left(\beta + 6\right)}{\beta (\beta + 1)} r^{-\beta}\ ,
\ee
which gives
\bea
\label{DMst19}
&& \omega(\phi)= \omega_0 \phi^{-\beta - 2}\ ,\quad
\omega_0\equiv - \frac{2\alpha\left\{-(\beta+1)^2 + 5\right\}}{\kappa^2(\beta + 1)} \ ,\nn
&& V(\phi) = V_0 \phi^{-\beta - 2}\ ,\quad
V_0\equiv \frac{\alpha(\beta^2 - 6)}{2\kappa^2 (\beta + 1)}\ .
\eea
Eq.(\ref{DMst18}) shows that, if the matter does not directly couple with
the scalar field $\phi$, the matter particle
receives the potential proportional to $r^{-\beta}$, that is, the force
whose strength is proportional to $r^{-1 - \beta}$.
As $0<\beta<1$, the force is stronger than the one produced by Newton
potential for large $r$. We should also note that, if $\alpha<0$, the
force is attractive. Hence, the above potential might explain the rotation
curves of the galaxies and the formation of
the galaxy clusters. It is remarkable that above potential contains the
powers of scalar in close analogy with cosmologically-viable potential
(\ref{STm13}).

We should also note that $\omega_0$ is positive if $\alpha$ is negative.
Therefore $\omega(\phi)$ is positive since
$\phi=r>0$. Then by using (\ref{STm4}), one finds
\be
\label{DMst20}
\varphi=-\frac{2\sqrt{\omega_0}}{\beta}\phi^{-\frac{\beta}{2}}\ ,
\ee
which gives the potential in (\ref{STm6}) as
\be
\label{DMst21}
V=V_0 \left( - \frac{\beta\varphi}{2\sqrt{\omega_0}}\right)^{2(1+2/\beta)}\ .
\ee
As $\phi$ is always positive, $\varphi$ is negative.
Thus, using method of refs.\cite{rec} (second section) it is demonstrated
that scalar-tensor theory with specific potential may produce the dark
matter effect. It is remarkable that the corresponding reconstruction
may be fulfilled for any requested modification of the Newton potential.

\section{Discussion}

\label{sec5}

In summary, we demonstrated that matter dominated era may be combined with
acceleration era within some class of scalar-tensor theory.
The explicit program of scalar-tensor theory reconstruction is presented.
  Clearly,
this makes non-realistic single scalar-tensor theory with other types of
potentials as dark energies due to appearence of cosmological bounds.
Nevertheless, the situation may be improved for several scalars (several
dark energy types) where the realization of sequence of matter dominated
and acceleration era should be investigated from the very beginning.
This will be studied in other place.

The same problem is discussed for dark energy fluid in terms of redshift
parametrization. It is shown that for some examples of EOS the matter
domination-dark energy transition may occur prior to
deceleration-acceleration transition. Hence, intermediate and late-time
stages of the universe may be unified also in frames of dark energy fluid.
As a by-product the reconstruction method which was used to get the
form of scalar potential from cosmological bounds is applied for the same
scalar-tensor theory to get the dark matter effect from it. The
scalar-tensor theory with requested modified Newton law at large scales is
constructed. Such a theory may explain the rotation curves of the galaxies
and the formation of galaxy clusters. It is interesting that the obtained
scalar potential seems to be qualitatively similar to the one obtained
from the cosmological bounds.

It seems that we are still at the beginning of the road to complete
  theory which unifies  all the known epochs being compatible
with astrophysical/cosmological data and Solar System tests.
Nevertheless, the fact that such unification is possible at least for
intermediate-time and late-time universe is quite promising.

\section*{Acknowledgements}

We thank S. Capozziello for helpful discussions.
The investigation by S.N. has been supported in part by the
Ministry of Education, Science, Sports and Culture of Japan under
grant no.18549001
and 21st Century COE Program of Nagoya University
provided by Japan Society for the Promotion of Science (15COEG01),
and that by S.D.O. has been supported in part by the project FIS2005-01181
(MEC, Spain), by the project 2005SGR00790 (AGAUR, Catalunya), by LRSS
project N4489.2006.02 and by RFBR grant 06-01-00609 (Russia).
H.S. acknowledges the support of the Secretar\'{\i}a de Estado
de Universidades e Investigaci\'{o}n of the Ministerio de Educaci\'{o}n
y Ciencia of Spain within the program ``Ayudas para movilidad de Profesores
de Universidad e Investigadores espa\~{n}oles y extranjeros"
and in part the support by MEC and FEDER under project 2004-04582-C02-01
and by the Dep. de Recerca de la Generalitat de Catalunya under contract
CIRIT GC 2001SGR-00065. H.S. would like to thank the Departament E.C.M.
of the Universitat de Barcelona for the hospitality.

\end{document}